\documentclass[aps, twocolumn,preprintnumbers, pra,floatfixt,10pt]{revtex4}
\pdfoutput=1
\usepackage{graphicx}
\usepackage{amssymb}
\usepackage{amsmath}

\usepackage{graphicx}
\usepackage{dcolumn}
\usepackage{bm}
\usepackage{graphicx}
\usepackage{amsmath, amsfonts, amssymb}
\usepackage{pstricks}
\usepackage{amsxtra}
\usepackage{amsthm}
\usepackage{verbatim}

\usepackage{setspace}

\newcommand{\ave}[1]{\langle#1\rangle}
\newcommand{\fref}[1]{Fig.~\ref{#1}}	
\newcommand{\hrho}{{\hat \rho}}
\newcommand{\eqnref}[1]{Eq.~(\ref{#1})}	
\newcommand{\abs}[1]{\left|#1\right|}
\newcommand{\secref}[1]{Section~\ref{#1}}	
\newcommand{\hvphi}{\hat\varphi}
\newcommand{\ket}[1]{\left|#1\right\rangle}
\newcommand{\adag}{{\hat{a}^\dag}}

\newcommand{\hc}{\mathrm{H.c.}}

\newcommand{\psidag}{\hat \psi^\dag}

\begin{document}
\title{Interference of parametrically driven one-dimensional ultracold gases}
\author{Susanne Pielawa}
\affiliation{Physics Department, Harvard University, Cambridge, MA 02138, USA}

\date{\today}
\begin{abstract}
We theoretically analyze interference patterns of parametrically driven one-dimensional ultracold atomic gases. By modulating the interaction strength periodically in time, we propose to excite collective modes in a pair of independent one-dimensional gases at energies corresponding to the drive frequency. The excited collective modes lead to spatial oscillations in the correlations of the interference pattern, which can be analyzed to obtain the sound velocity of the collective modes. We discuss both bosonic and fermionic systems, and how such experiments could be used to probe spin charge separation. 
\end{abstract}
\maketitle


\section{introduction}

One-dimensional systems show many interesting properties like charge fractionalization and separation of charge excitations from spin excitations. There is 
experimental evidence for these properties in various systems \cite{Y.Jompol07312009,PhysRevLett.77.4054,KimNaturePhysics,SegoviaObervationSpinCharge,PhysRevB.68.125312,TunnelingSpectroscopy_1D_Wire,AuslaenderSpinChargeSep}. 
Recent progress in trapping and cooling of cold atoms, 
has made it possible to study one-dimensional systems and their properties in both fermions and bosons~\cite{PhysRevLett.87.160405,
PhysRevLett.91.250402,
SpinChargeSeparationBosons, ToshiyaKinoshita08202004,ExpTonksGirardeauGas,Recati1DFermi}. One of the primary tools in measuring properties of cold atoms are interference experiments.

Surprisingly, it has been shown that interfering two independently created Bose Einstein condensates (e.g. by releasing the confining potential and letting them expand and overlap) gives rise to spatially periodic patterns in each experimental shot. As there is no well defined phase between the two condensates, the maxima and minima of the pattern are at different positions in each shot, so that the average over many shots, corresponding to a quantum-mechanical ensemble average $\ave{\hat \rho (\vec r)}=0$, contains no interference fringes. 
The key to this effect is that we can decompose the state of the two condensates, each containing a definite number of particles, into a superposition of states of well defined relative phases. Measuring the interference pattern collapses the wavefunction of the combined system to a state with definite phase, and thus produces a periodic pattern. 
To theoretically study this effect, we need to study the density-density correlation function $\ave{\hat\rho(\vec r_1)\hat\rho(\vec r_2)}$ instead of the average density, since the quantum mechanical average of the density formed in interference experiments samples all possible relative phases and thus hides the interference pattern. 
The above arguments hold true even for an interfering pair of one-dimensional systems, with one important modification: as there is no long range order, the maxima formed in the interference pattern are wavy lines instead of straight lines. This waviness contains information about the correlation function within each 1d system \cite{Hofferberth, imambekov-2007}. 

In this paper, we propose an experimental approach to create and directly probe excitations in one-dimensional ultracold atom systems of both bosons and fermions. 
In particular, for systems composed of mixtures of two species, our method allows us to see separation to spin and charge modes. Our approach is to start with a pair of one-dimensional systems, parametrically drive excitations by temporally periodic modulation of the particle particle interaction strength, and finally analyze the interference pattern. 
%
Experiments involving parametric driving to create excitations in one-dimensional fermionic gases have recently been proposed in Refs \cite{kagan:023625, graf-epl-2009}. 
In this work, in addition to fermionic gases, we study also bosonic gases, which are of experimental interest.
%

\begin{figure}[bt]
\includegraphics[width=0.5\textwidth]{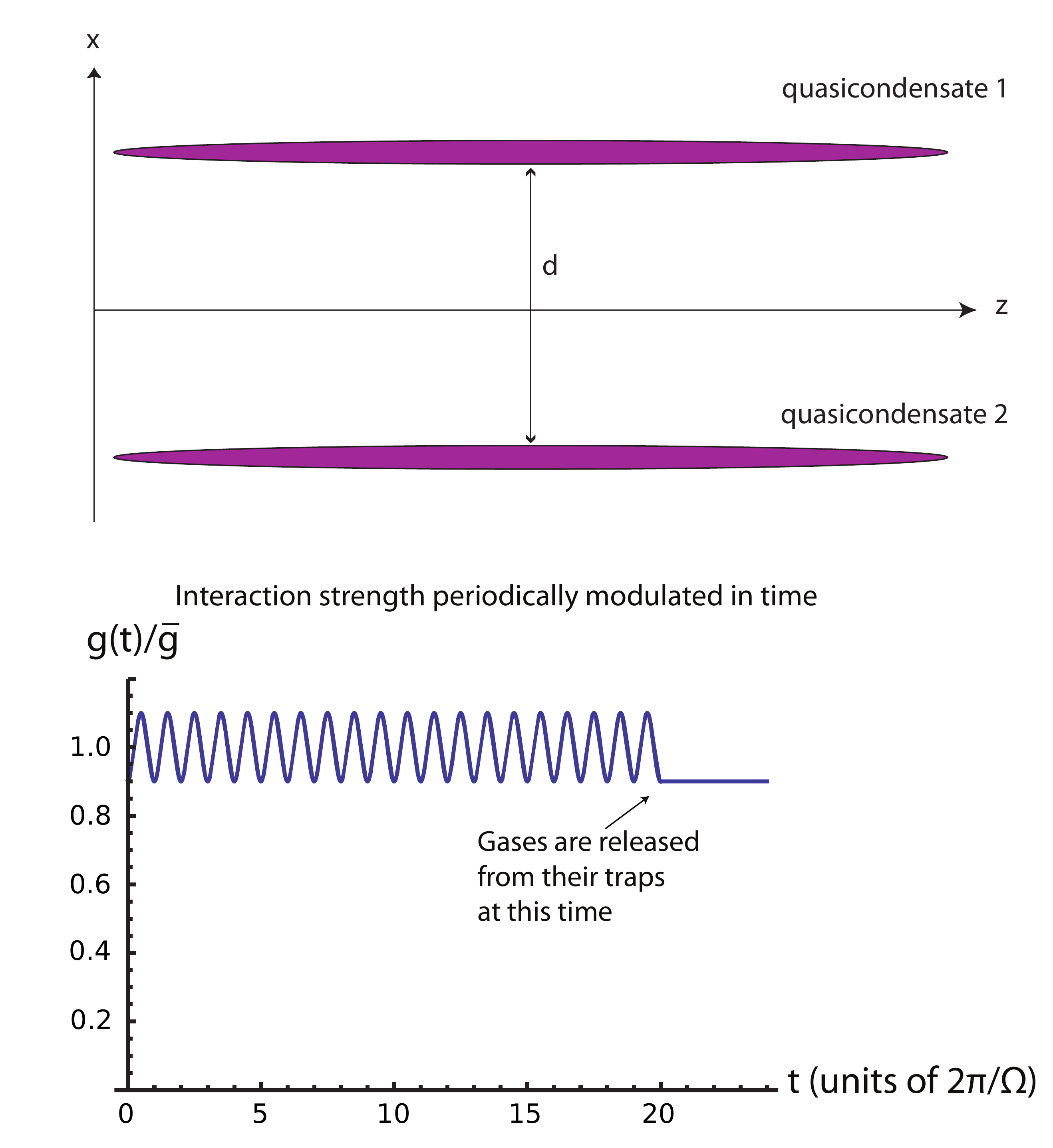}
\caption{(Color online) 
Proposed setup: two independently created one-dimensional systems of ultracold atoms are first parametrically driven by changing the inter-atomic interaction strength periodically in time (e.g. by changing the transverse trapping frequency), at time T the driving stops and the systems are released from their traps and allowed to overlap.  
An absorption image is taken, showing wiggely fringes along the z axis. These fringes contain information about  the two-point correlation function within each gas (before expansion) and can be analyzed to obtain the sound velocity of the collective modes. 
}
\label{fig1}
\end{figure}

In interference experiments with one-dimensional systems of cold atoms, the gases are released from their traps and allowed to expand. We label the longitudinal direction as the $z$ the axis, and the two systems lie in the $x$-$z$ plane, see \fref{fig1}. Due to the strong transverse confinement, when released from their traps, the atoms expand mainly in the transverse directions. Therefore, we shall neglect the motion of atoms in the longitudinal direction during the expansion. Once the gases have expanded to a size much larger than the initial distance $d$ between the two systems, an absorption image is taken. We neglect the slowly varying envelope and the expectation value of the density in case of independently created quasicondensates is a constant, $\ave{\hrho(x,z)}=\rho_{\rm{tof}}\rho_0$, where $\rho_0$ is the line density of each quasicondensate. The density-density correlations function, however, contains an oscillating term at wave vector $Q=\frac{dm}{\hbar t}$, $m$ is the particle mass, $t$ the time of expansion:
\begin{equation}
\ave{\hrho(r_1)\hrho(r_2)}\propto\left( \rho_0^2\pm \left|\ave{\psidag(z_1)\hat \psi(z_2)}\right|^2 \cos(\Delta x Q)\right),
\label{eq1}
\end{equation}
where $r_i$ stands for $(x_i,z_i)$, $\Delta x = x_1-x_2$, the upper (lower) sign applies to bosons (fermions). $\hat\psi(z)$ denotes the particle annihilation operator in either of the equivalent initial 1d systems. To derive \eqnref{eq1} we have neglected the motion of atoms along the longitudinal direction during expansion; assumed that the two systems are initially identical and independent; assumed large expansion times (e.g. the size of the overlapping clouds is much larger than their initial separation $d$). See appendix \ref{appendix1}; Ref~\cite{imambekov-2007} contains a more detailed discussion. 

Typically, experimental data is analyzed by looking at the the integrated density-density correlation function~\cite{Hofferberth, Polkovnikov04182006},
\begin{equation}
\int_0^L  dz_1 dz_2 \ave{\hrho(x,z_1)\hrho(0,z_2)} \propto \rho_0^2 L^2+ \ave{\left| A_L\right|^2}\cos(Qx), \nonumber
\end{equation}
where $L$ is the length of the integration, and the interference amplitude $A_L$ is given by 
$\ave{\left| A_L\right|^2}=L\int_0^L dz \left|\ave{\psidag(z)\hat \psi(0)} \right|^2$. 
One can now define the interference contrast $C_L=\abs{A_L}/\rho_0 L$, 
which decays as the integration length $L$ is increased, encoding the decay of the correlation functions
in the one-dimensional systems \cite{Hofferberth}. 

Our approach is to study the interference pattern of one-dimensional gases that have been parametrically driven out of equilibrium. The driving is done by a temporally periodic change of interaction strength 
for a certain time $T$ before the gases are released from their traps. 
To extract information from the interference pattern, we suggest to take the Fourier transform along the z-axis, and then study correlations for that Fourier transform. We define
\begin{equation}
\hat{\tilde\rho}(x,q)=\int_0^L e^{iqz}\hrho(x,z)dz
\end{equation}
and then consider the quantity $\ave{\hat{\tilde\rho}(x,q)\hat{\tilde\rho}(0,-q)}$ which is given by
\begin{eqnarray}
\ave{\hat{\tilde\rho}(x, q)\hat{\tilde\rho}(0,-q)}
&\propto&\cos(xQ)L\int_{0}^{L} dz \cos(qz) \abs{\ave{\hat\psi^\dag(z)\hat\psi(0)}}^2.\nonumber
\end{eqnarray}
Thus, by Fourier transforming the interference pattern, one can obtain the Fourier transform of the correlation function of a single 1D gas,
\begin{equation}
\abs{A(q, T)}^2:=L\int_{0}^{L} dz \cos(qz) \abs{\ave{\hat\psi^\dag(z)\hat\psi(0)}}^2.
\end{equation}
The quantity $\abs{A(q, T)}^2$ is also a function of the total driving time $T$ (we have suppressed this time argument $T$ in $\hat \psi$ and $\hat \rho$ in the above equations). 
This analysis can in principle be done for fermions and bosons.

Our main results are: (1) We find that parametric drive at frequency $\Omega$ creates pairs of excitations with total momentum zero. Therefore, the quantity $\abs{A(q)}$ 
shows a resonance peak at the wave vector corresponding to half the driving frequency, $q=\Omega/2 u_i$ (and, in principle, associated multiples), where $u_i$ is the sound velocity of the corresponding mode. Such an analysis can be used to find the sound velocity of the 1D bose system. 
(2) For a two component Bose system, we find two primary peaks corresponding to ``spin" and ``charge" velocity. (3) Similarly one could analyze an interference pattern of a driven 1D system of two component fermions, where one expects two peaks corresponding to spin and charge excitations. For fermions, however, there are some complications due to the rapid decay of the correlation functions, which we will address in \secref{SpinCharge}.

The paper is organized as follows: In \secref{DrivingSqueezing} we introduce the formalism and describe the relation between a driven 1d gas and a driven simple harmonic oscillator. In \secref{Detecting} we calculate the correlation functions of driven systems and show how information about the driving can be extracted from the interference pattern. In \secref{SpinCharge} we address the problem of observing spin charge separation for fermionic 1d systems. 

\section{Parametric driving and Luttinger mode squeezing}
\label{DrivingSqueezing}

\subsection{Description of a 1d system: Luttinger liquid}
One-dimensional systems of interacting bosons or fermions can in general be described by Hamiltonian
\begin{eqnarray}
\hat H &=& \int_0^L -\frac {\hbar^2} {2m} \psidag_\sigma(z) \nabla^2 \hat \psi_\sigma(z)\\
&&+\int_0^L\int_0^L g_{\sigma\sigma^\prime}(z-z^\prime)\psidag_\sigma(z)\hat\psi_\sigma(z)\psidag_{\sigma^\prime}(z^\prime)\hat\psi_{\sigma^\prime}(z^\prime) dz dz^\prime,
\nonumber
\end{eqnarray}
where $g(z)$ is the interaction strength, and $\psidag(z)$ creates a particle at point $z$. Low energy excitations may in turn be described by the Luttinger liquid Hamiltonian which in momentum space is given by a sum of harmonic oscillators
\begin{equation}
\hat H = \frac 1 2 \sum_{q\ne 0} \left(
\frac{\pi u}{K} \hat \Pi_q \hat \Pi_{-q} + \frac{u K }{\pi} q^2 \hvphi_q\hvphi_{-q}
\right).
\label{LL_mom_space}
\end{equation}
where $\hat\Pi$ and $\hat\varphi$ are conjugate fields: $\left[\hat\varphi_q, \hat\Pi_{q^\prime}\right]=i\delta_{q, -q^\prime}$, 
$K$ and $u$ are the Luttinger parameters, $u$ corresponds to the sound velocity, and $K$ is a dimensionless parameter. 
For a finite system of length ${\cal L}$ we quantize the momentum $q=\frac{2\pi}{\cal L}n$. 
We shall first consider the case of spinless bosons and then generalize to a two component Bose mixture and fermions with spin. 
For a weakly interacting one component Bose system ($K>1$) the Luttinger parameters are given by
\begin{equation}
K=\pi\hbar\sqrt{\rho_0/gm},\qquad\qquad
u=\sqrt{\frac{\rho_0 g}{m}}
\label{eq:LuttingerParameters}
\end{equation}
where $\rho_0$ is the 1D line density, and g is the interaction parameter, related to the three dimensional scattering length $a_s$ through $g=2h\nu_T a_s$~\cite{Hofferberth_nature, cazalilla-2004-37}, where $\nu_T$ is the transverse trapping frequency. The relation between the physical boson creation operator and the Luttinger fields is given by
\begin{equation}
\psidag(z) = \sqrt{\rho_0+\hat \Pi (z)}e^{-i\hvphi(z)}
\end{equation}
and the correlation function in the ground state has a power law decay
\begin{equation}
\ave{\psi^\dag(z)\psi(0)}_0 \propto \rho_0 \left(\frac  {z}{\xi_h}\right)^{-\frac {1} {2K}}, 
\end{equation}
where $\xi_h=2K/\rho_0$ is a short distance cutoff. 
Let us now assume that there are two species of bosons (e.g. two hyperfine states or two different atoms), labelled $\uparrow$ and $\downarrow$. We further assume that the interaction between two atoms of the species labelled up and the one between two atoms of the species down are the same, $g_{\uparrow \uparrow}=g_{\downarrow \downarrow}=g_\parallel$, 
a condition that can be realized in experiments using hyperfine states $\ket{F=1,m_F=-1}$ and $\ket{F=2,m_F=+1}$ of $^{87}$Rb \cite{widera:140401,Hofferberth,PhysRevA.69.032705}. 
We label $g_\perp$ the interaction strength between different species. In this case the Hamiltonian separates into a ``charge" part with $g_c = g_\parallel+g_\perp$ and ``spin" part with $g_s=g_\parallel-g_\perp$. The ground state correlation function becomes
\begin{equation}
\ave{\psi^\dag(z)\psi(0)}_0 \propto \rho_0 \left(\frac  {z}{\xi_h}\right)^{-\frac {1} {4 K_c}-\frac {1} {4 K_s}}
\end{equation}
where $K_c$ ($K_s$) is related to $g_c$ ($g_s$) through \eqnref{eq:LuttingerParameters}.\\

Next let us consider a fermionic system of two components labelled $\uparrow$ and $\downarrow$. In the case of ultracold atomic gases, two fermions of the same species cannot interact via s-wave scattering ($g_\parallel=0$). In the following we assume $\delta$-interactions and denote by $g$ the interaction constant between two fermions of opposite spin. The Hamiltonian separates in a charge part $H_\rho$ and a spin part $\hat H_\sigma$. Both charge and spin part have the form of \eqnref{LL_mom_space} with the Luttinger parameters given by~\cite{Giamarchi}
\begin{subequations}
\begin{eqnarray}
u_\nu K_\nu &=& v_F,\\
\frac{u_\nu} {K_\nu} &=& v_F \left(1\pm  \frac{g}{\pi v_F} \right),\label{eqn:uFermi}
\end{eqnarray}
\end{subequations}
here $\nu= \rho, \sigma$ and the upper (lower) sign applies to the charge (spin) mode. 
The single particle annihilation operators are given by $\psi_\uparrow(z)=\psi_{\uparrow, R}(z)+\psi_{\uparrow, L}(z)$, where the index $R$ ($L$) stand for a right (left) moving particle, and 
\begin{subequations}
\begin{eqnarray}\hat\psi_{\uparrow R}^\dag(z)&\rightarrow&\sqrt{\rho_0} e^{-i k_F z} e^{-i(-\hvphi_{\uparrow}(z)+\theta_{\uparrow}(z))},\\
\hat\psi_{\uparrow L}^\dag(z)&\rightarrow&\sqrt{\rho_0} e^{i k_F z} e^{-i(\hvphi_{\uparrow}(z)+\theta_{\uparrow}(z))},
\end{eqnarray}
\label{eqn:fermipoerators}
\end{subequations}
analogously for $\downarrow$. The fields which decouple the Hamiltonians in $H_\rho$ and $H_\sigma$ are given by
\begin{eqnarray}
\varphi_\rho&=&\frac 1 {\sqrt 2} \left(\varphi_\uparrow+\varphi_\downarrow \right),\\
\varphi_\sigma&=&\frac 1 {\sqrt 2} \left(\varphi_\uparrow-\varphi_\downarrow \right).
\end{eqnarray}
and same for $\theta_\nu$, which enters in the Luttinger liquid Hamiltonian as $\nabla \theta_\nu(z) = \pi \Pi_\nu(z)$.

The spin part of the Hamiltionan has an additional sine Gordon term, arising from backscattering between fermions with opposite spin ($g_{1\perp}$ term). However, this term renormalizes to zero~\cite{Giamarchi} if $K_\sigma>1$ (repulsive interactions),
otherwise it creates a gap in the spectrum of the spin Hamiltonan. In this paper we concentrate on repulsive interactions and the sine-Gordon term is not relevant. 

 The single particle correlation function in the ground state
 \begin{equation}
 \ave{\psi_\uparrow(z) \psi_\uparrow(0)}
 =2\rho_0\cos(k_F z)  \left( \frac z {\xi_h}
 \right)^{-\frac 1 4\left(K_\rho+\frac 1 {K_\rho}+K_\sigma+\frac 1 {K_\sigma}\right)}
 \end{equation}
 is oscillating at the length scale of particle spacing. 

\subsection{Parametric driving and squeezing}%

\begin{figure*}
\begin{tabular}{l l}
a)&b)\\
\begin{minipage}{0.49\textwidth}
\includegraphics[width=.95\textwidth]{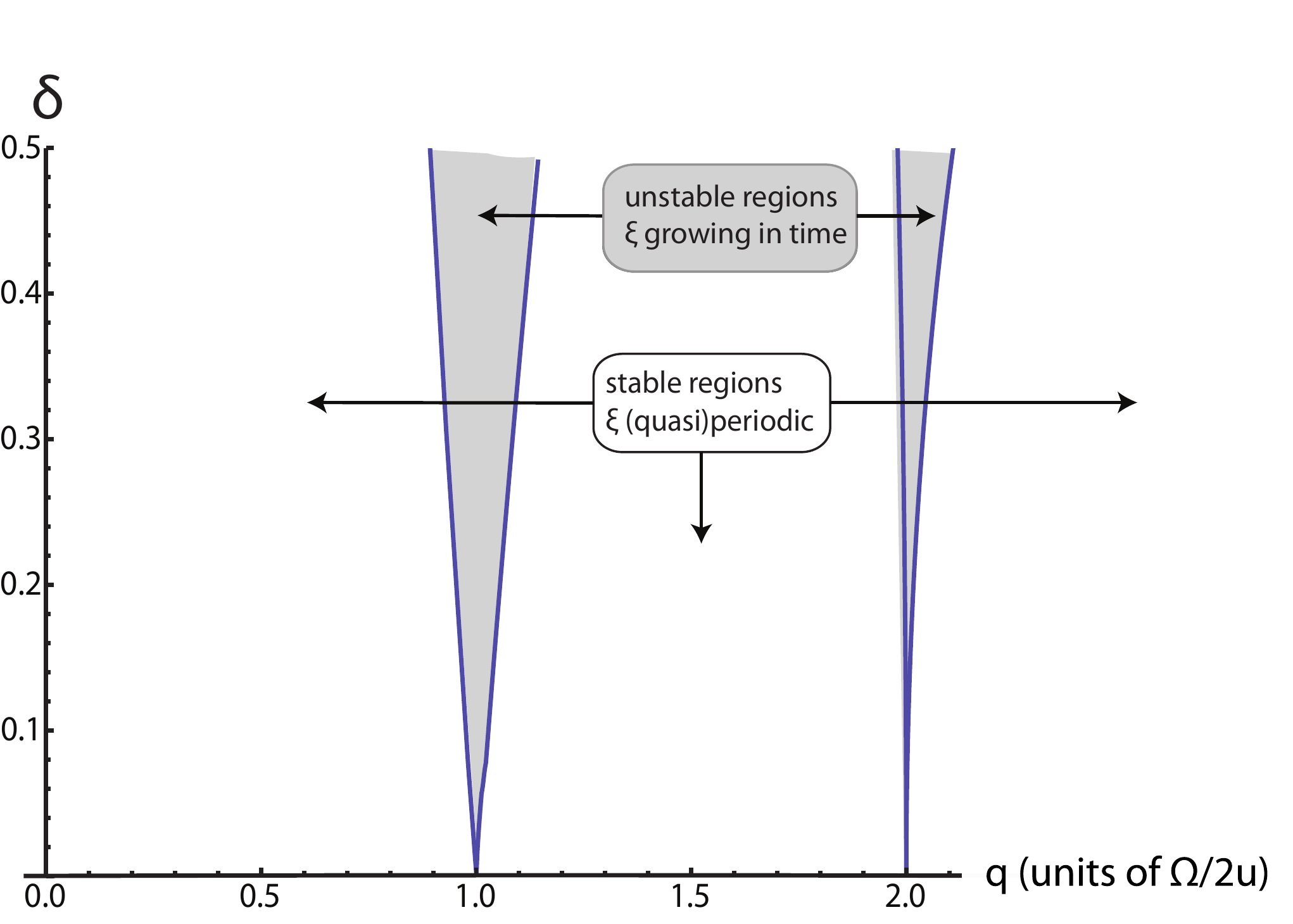}
\end{minipage}
&
\begin{minipage}{0.49\textwidth}
\includegraphics[width=.9\textwidth]{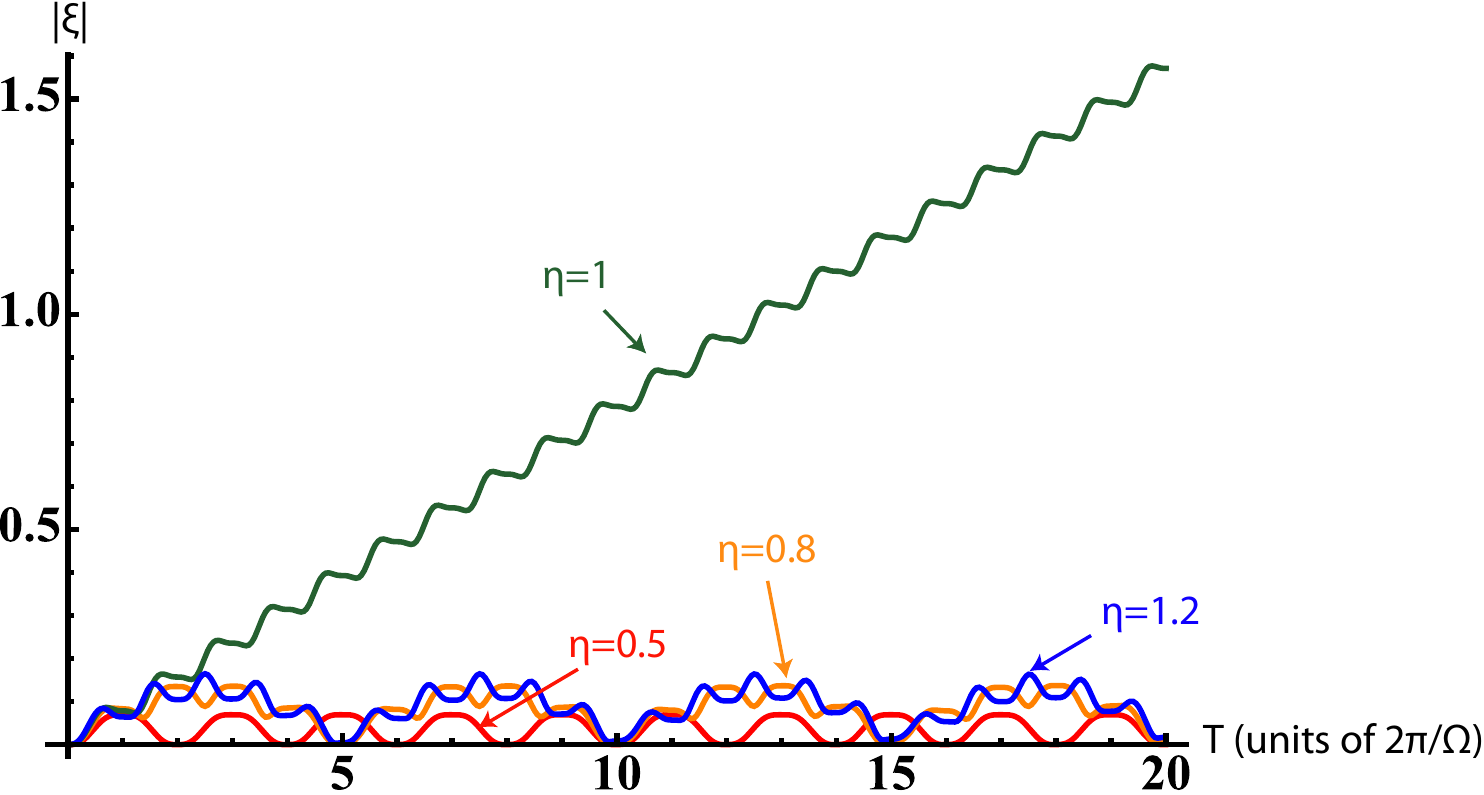}
\end{minipage}
\\
\end{tabular}
\caption{(Color online) a) Stability diagram of \eqnref{eq:Mathieu}, as a function of momentum q (in units of $\frac {\Omega} {2\bar u}$) and relative driving amplitude $\delta$.  
Gray shaded regions are resonant, where the number of excitations in that mode grows exponentially in time. White regions are off-resonant and the number of excitations is a (quasi-) periodic function in time and thus bounded from above. 
b) Absolute value of squeezing parameter $\abs{\xi}$ as a function of time for different values of $\eta=\frac{2\omega_q}{\Omega}$, at fixed $\delta=.1$. In the unstable region, $\xi$ grows linearly in time. 
The first resonance occurs at $2 u q = \Omega$, when the driving frequency is twice the frequency of an excitation in that mode, since excitations are created in pairs $(-q, q)$.
}
\label{fig:stability}
\end{figure*}
Experimentally the interaction parameter $g$ can be changed by either changing the scattering length $a_s$, or by changing the transverse confinement. 
Let us now imagine a 1d system that is parametrically driven by changing $g$ periodically in time for a certain time $T$,
\begin{equation}
g(t) = \bar{g} \left(1-\delta^\prime \cos(\Omega t)\right),
\end{equation}
where $\bar{g}$ is the average value of interaction and $\Omega$ is the driving frequency.
Since the Luttinger Hamiltonian \eqnref{LL_mom_space} is just a collection of harmonic oscillators, this situation can be mapped to a collection of parametrically driven simple harmonic oscillators via $\hat P_q = -\sqrt{uKq^2/\pi} \hat \varphi_q$ and 
$\hat Q_q=\sqrt{\pi/uKq^2} \hat \Pi_q$. We note that this mapping is time-independent for both the bosonic and the fermionic case, as $uK$ remains constant.
Each pair of modes $(q, -q)$ is then described by 
\begin{equation}
\hat H_q(t) = \frac 1 2 \hat P_q \hat P_{-q} + \frac {\omega_q^2} 2 \left[ 1-\delta \cos(\Omega t)\right] \hat Q_q \hat Q_{-q}
\end{equation}
with 
\begin{equation}\omega_q=\bar u q,
\end{equation}
and $\bar u^2=\bar g \rho_0$, $\delta=\delta^\prime$ for bosons and $\bar u_{\nu}^2=v_F^2(1\pm \frac {\bar g}{\pi v_F})$, $\delta=\pm \delta^\prime \frac{\bar g}{\pi v_F\pm \bar g}$ for fermions. 
It is well know that a parametrically driven quantum harmonic oscillator can be mapped to a static reference system~\cite{PhysRevA.51.950}. If initially in the ground state, the final state after a driving time $T$ is, up to an overall phase factor, given by
$\hat S\left(\xi_q(T)\right) \ket 0$
where 
\begin{equation}
\hat S(\xi_q(T))= \exp \left[\xi^*_q(T) \hat a_q \hat a_{-q}- \xi_q(T) \adag_q \adag_{-q} \right]
\end{equation}
is the squeezing operator, and $\hat {a}_q$ ($ \adag_q$) is the annihilation (creation) operator of one excitation in the static reference system, 
\begin{equation}
\hat a_q = \sqrt{\frac {\omega_{\rm ref}(q)} 2}\left(\hat Q_q + \frac i{\omega_{\rm ref}(q)} \hat P_q\right).
\end{equation}
We choose our reference system to have the eigenfrequency of the system at $t=0$, $\omega_{\rm ref}(q)=\omega_q\sqrt{1-\delta}$.
\newcommand{\x}{{\mathcal X}} \newcommand{\y}{{\mathcal Y}}
The squeezing parameter $\xi_q(T)$ can be expressed in terms of solutions of the classical parametrically driven harmonic oscillator, namely solutions to the following second order differential equation
\begin{equation}
z^{\prime\prime}(\tau)+\left(\frac{2\omega_{q}}{ \Omega}\right)^2 \left[1 - \delta \cos(2\tau)\right]z(\tau) = 0,
\label{eq:Mathieu}%
\end{equation}%
\newcommand{\etar}{\eta_{\rm ref}}%
which we have expressed in dimensionless variables, and $\tau=\frac{\Omega t}{2}$. We define $\eta=\frac{2\omega{q}}{ \Omega}$, and $\etar=\eta\sqrt{1-\delta}$.
Let ${\mathcal X}(\tau)$ and ${\mathcal Y}(\tau)$ be solutions of this equation with $\x(0)=\y^\prime(0)=1$ and $\x^\prime(0)=\y(0)=0$, and define $z(\tau)=\x(\tau)+i\etar\y(\tau)$, then $\xi_q$ is defined through
\begin{equation}
4 \cosh^2(\abs{\xi_q}) = \abs{z}^2+\abs{z^\prime}^2/\etar^2 +2
\end{equation}
and an expression for its phase, $\xi = \abs{\xi} e^{i \vartheta}$ , can be found through $\vartheta = \vartheta_u + \vartheta_v $, where $\abs{u} = \cosh\abs{\xi}$, and~\cite{PhysRevA.51.950}
\begin{subequations}
\begin{eqnarray}
&&u = \abs u e^{i \vartheta_u}=\frac {e^{-i \omega_r t}} 2 \left[ z(t) - \frac {i \dot z (t)} {\omega_r}\right],\\
&&v = \abs v e^{i \vartheta_v}=\frac {e^{i \omega_r t}} 2 \left[ z(t) + \frac {i \dot z (t)} {\omega_r}\right].
\end{eqnarray}%
\label{eq:squeezingparam}%
\end{subequations}%
Physically relevant quantities can be expressed in terms of the above, e.g. average number of excitations $\ave{\adag\hat a} = \sinh^2\abs{\xi}$, and uncertainties in position and momentum variables: $\etar\frac \Omega 2 \Delta\hat x^2 = \abs{z}/2$ and $\etar\frac 2 \Omega \Delta p^2= \abs{z^\prime}/2$. 

\begin{figure*}[t!]
\begin{tabular}{l l}
a)&b)\\
\begin{minipage}{0.49\textwidth}
\includegraphics[width=\textwidth]{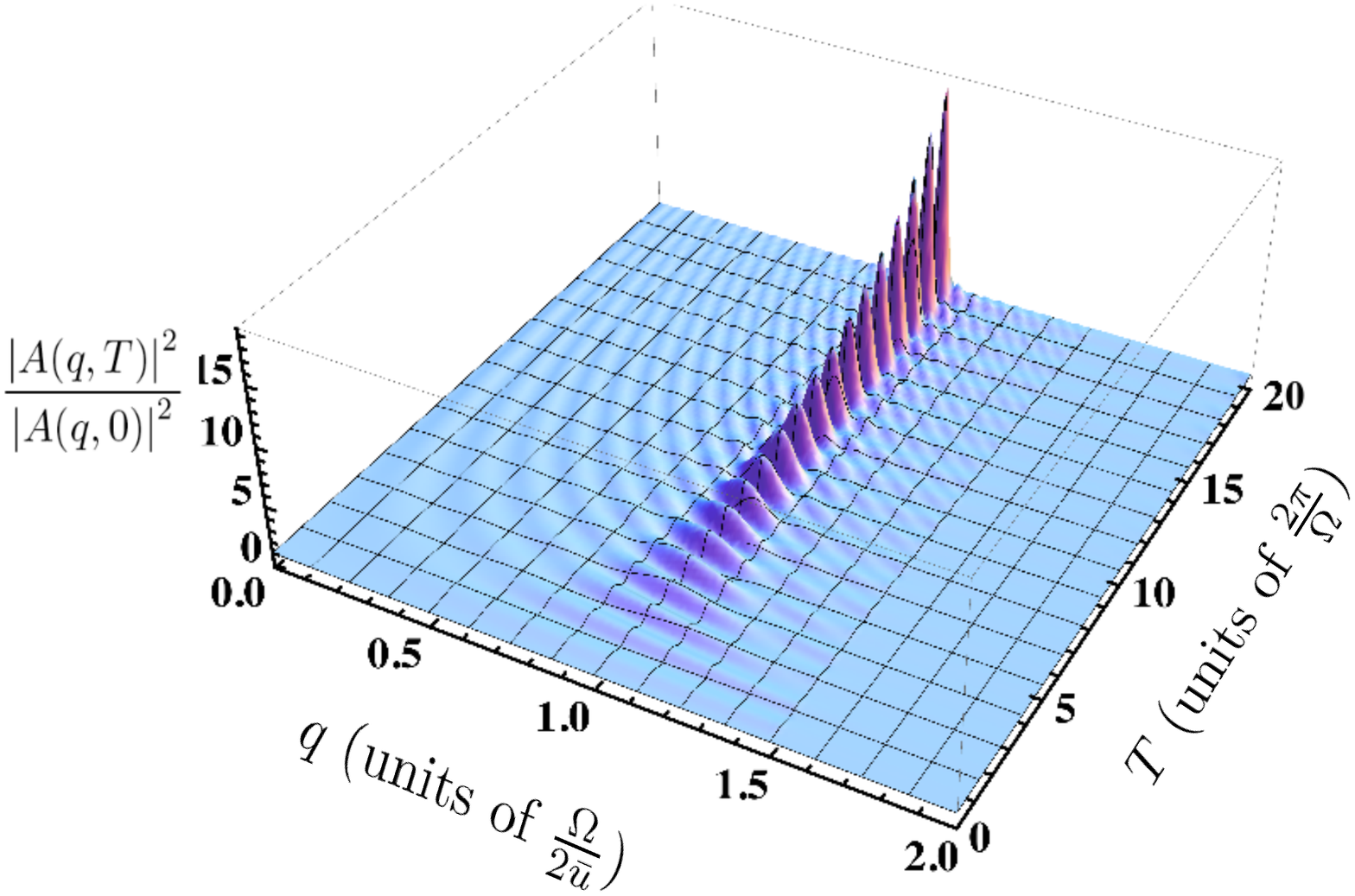}
\end{minipage}
&
\begin{minipage}{0.49\textwidth}
\includegraphics[width=.9\textwidth]{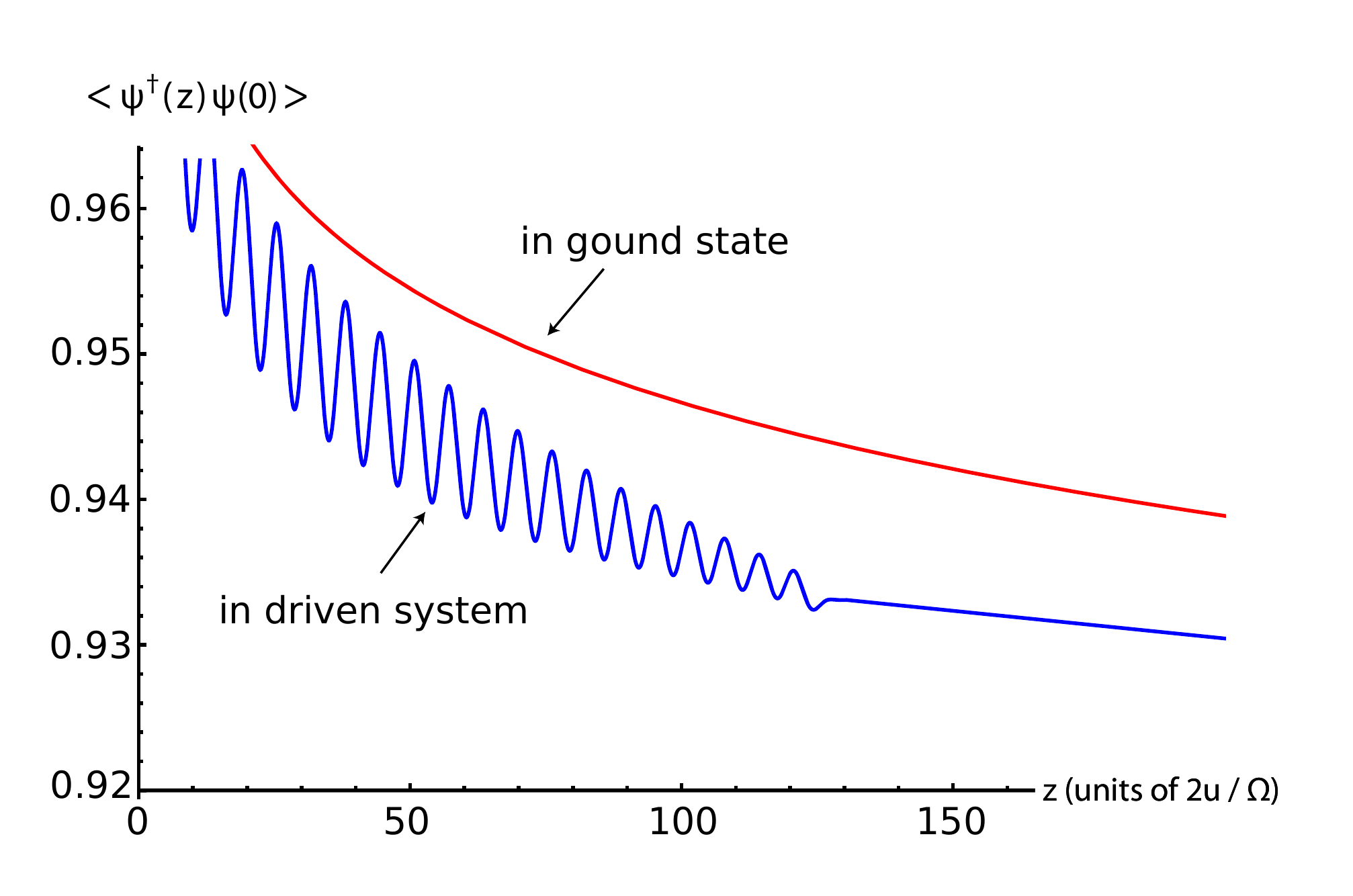}
\end{minipage}
\\
\end{tabular}
\caption{(Color online) 
Driven one-dimensional Bose gas: (a) Observable quantity $\ave{\abs{A(q,T)}^2}$, divided by its ground state value,  as a function of wavevector $q$ and number of oscillations complete during driving time $T$. At $T=0$ the system is in the ground state. 
As driving continues a resonance peak corresponding to the drive frequency and sound velocity emerges and grows in time. The relative driving amplitude here is $\delta=.1$. (b) Correlation function after driving time $T=20 \frac{2\pi}{\Omega}$. 
}
\label{fig:bosons}
\end{figure*}

We note that \eqnref{eq:Mathieu} is Mathieu's equation and its odd and even solutions $\x(\tau)$ and $\y(\tau)$ are referred to as Mathieu S and Matheiu C functions, because of their resemblance to sinusoidal functions in certain parameter regimes. Depending on parameters $\eta$ and $\delta$, this equation has stable, i.e. (quasi-)periodic solutions or unstable i.e. exponentially growing and decaying solutions, see \fref{fig:stability}. On resonance $\eta\approx 1$ the squeezing parameter grows as $\abs{\xi}\approx\frac \delta 8 T \Omega$. 

To summarize the above, the final state of a parametrically driven 1d system is fully described by the set of squeezing parameters $\xi_q$, which in turn is given by solutions of \eqnref{eq:Mathieu}. 

\section{Detecting excitations via interference}%
\label{Detecting}

We now show how the excitations of the driven condensate change the correlation function and the interference pattern. While the parameter $K$ can be extracted from the ground state interference pattern, interference of driven systems could be used to measure the sound velocity $u$.  
Given a set of squeezing parameters $\xi_q$ we calculate the correlation function for bosons and fermions and obtain $\ave{\abs{A_q}^2}$, a quantity that can be extracted from the interference pattern in experiments.   

\subsection{Bosons: one and two component systems}

The correlation function 
can be written in terms of the quadratures $\ave{\hvphi_q\hvphi_{-q}}_t$ as
\begin{eqnarray}
\ave{\psidag(z)\hat \psi(0)}&=&\rho_0\left(\frac {z}{\xi_h}\right)^{-\frac 1 {2K}} e^{-\frac 1 K \int d(\eta) \sin^2\left(\frac{\eta \tilde z}{2}\right)d\eta}
\label{eq:corrDriven}
\end{eqnarray}
where $\tilde z = \frac {\Omega}{2u} z$ is the normalized dimensionless position variable. The function $d(\eta)$ is defined as the change in quadrature $\ave{\hvphi_q\hvphi_{-q}}_t$ due to the excitations, 
\begin{eqnarray}
d\left(\frac {2\bar u} \Omega q\right)&:=&\frac {K \Omega} {2 \bar u} \left(\ave{\hvphi_q\hvphi_{-q}}_t-\ave{\hvphi_q\hvphi_{-q}}_0\right)
\end{eqnarray}
and can be expressed in terms of the squeezing parameters $\xi_\eta=r_\eta e^{i\vartheta_\eta}$
\begin{equation}
d\left(\frac {2\bar u} \Omega q\right)=\frac 1 q \left(\cosh(2\abs{\xi_q})-\cos\vartheta_q \sinh(2\abs{\xi_q})-1\right).\nonumber
\end{equation}
We calculate the correlation function and $\ave{\abs{A_q}^2}$ of a 1D bose gas after it has been parametrically driven. The integral in the exponent of \eqnref{eq:corrDriven} is evaluated numerically. \fref{fig:bosons} shows a result, as expected there is a peak in $|A_q|^2$  at a momentum corresponding to the driving frequency. 

\begin{figure*}
\begin{tabular}{l l}
a) & b)\\
\begin{minipage}{0.49\textwidth}
\includegraphics[width=\textwidth]{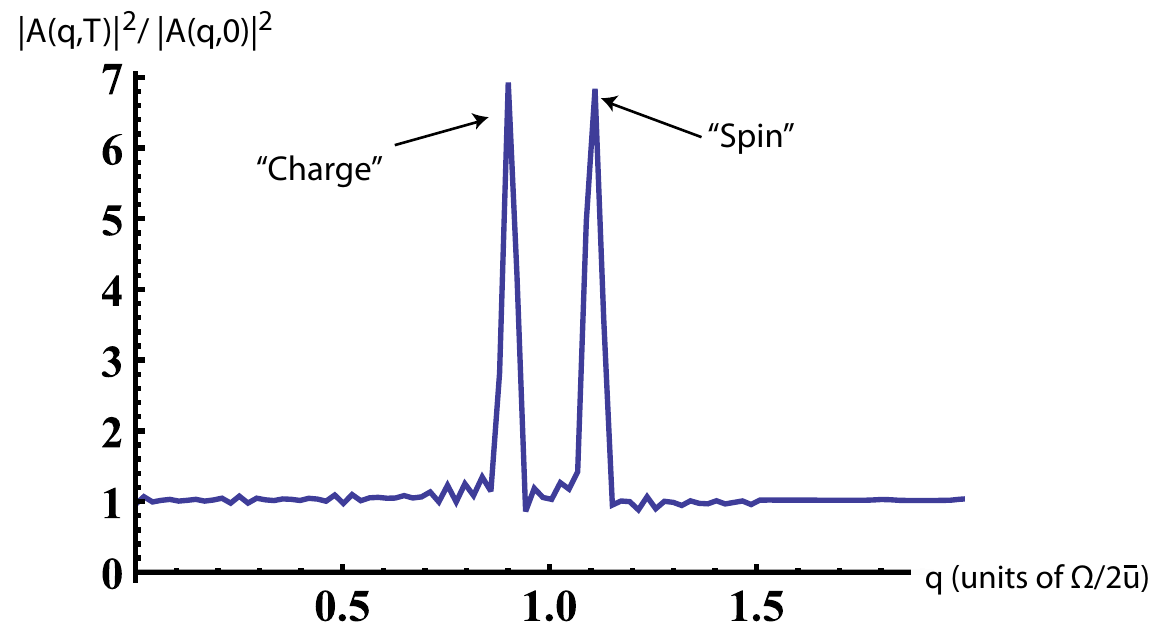}
\end{minipage}
&
\begin{minipage}{0.49\textwidth}
\includegraphics[width=.8\textwidth]{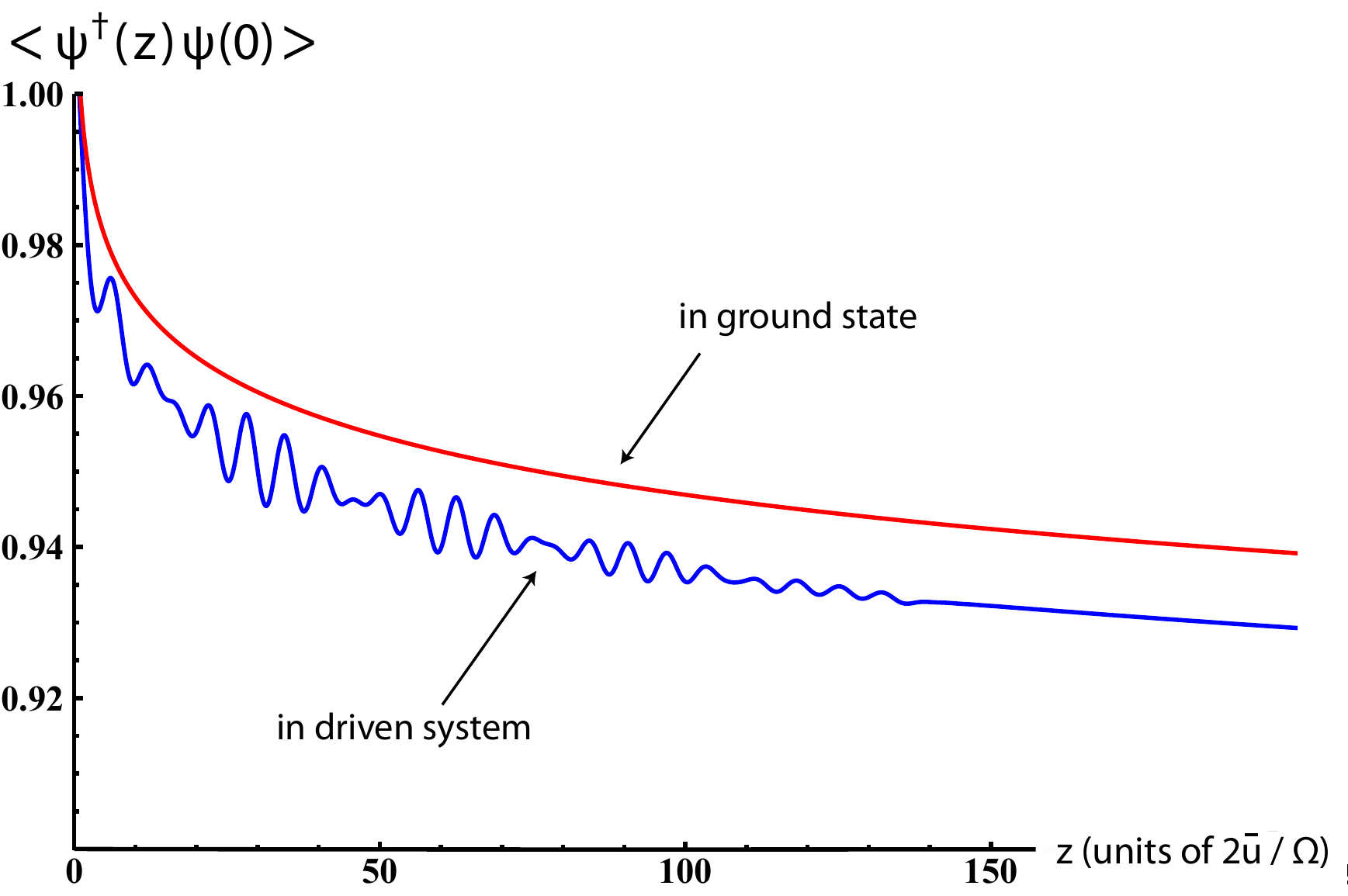}
\end{minipage}
\\
\end{tabular}
\caption{(Color online) Observable quantity $\ave{\abs{A_q}^2}$ and correlations function for parametrically driven two component one-dimensional Bose gas. The parameters here chosen are $\delta=.1$ and number of oscillations in driving 
$n = \left(\frac{T\Omega}{2\pi} \right) =20$. 
}
\label{fig:2comp_bosons}
\end{figure*}

There are two modes in a two component bose system with different sound velocities. The parametric driving excited each of the modes independently. This creates beats in the correlation function, which can be seen as peaks in $A_q$: there are then two peaks, one corresponding to each mode, see \fref{fig:2comp_bosons}. This setup can be used to probe spin-charge separation in continuous bosonic systems. It has previously been proposed to use continuous bosonic one-dimensional systems to probe spin-charge separation~\cite{SpinChargeSeparationBosons} . 

\subsection{Two component Fermi systems}
We first calculate the correlation function for the driven fermi gas. Note that both fields appear in the exponent of single particle operators, see \eqnref{eqn:fermipoerators}. We find for each mode 
\begin{subequations}
\begin{eqnarray}
\ave{\psidag_R(z) \hat \psi_R(0)} &=& \rho_0 e^{-ik_F z}  \left(\frac{z}{\xi_h} \right)^{-\frac 1 2 \left(K+\frac 1 K\right)}e^{-I(z)}\\
I(z)&=&\int d\eta \sin^2\left(\frac{\eta \tilde z}{2}\right)\left[\left(K+\frac 1 K\right) \sinh^2(r_\eta)-\right.\nonumber\\
&&\qquad \left.\abs{K-\frac 1 K} \cos\vartheta_\eta \sinh(2 r_\eta)\right].
\end{eqnarray}%
\label{eqn:corrfunctdrivenFermions}%
\end{subequations}%
The above formula is valid for spin-polarized fermions, and also for spin and charge part of a two component system of fermions. In that case, we just have to replace the constants (e.g. interaction strength $g$ and what follows from it) by the corresponding value for the spin- or charge part. \\
As a first approximation, let us neglect the second part of $I(z)$ with respect to the first part. There are two reasons for that: first, for fermions the Luttinger parameter $K$ is close to one (in fact, for the spin part it renormalizes to $K_\sigma=1$),  second, only the second term contains the phase of the squeezing parameter $\cos \vartheta_q$, which will average out if the systems are left to evolve after driving and before interfering. We then have
\begin{equation}
I(z)\approx \textrm{const}- \left(K + \frac 1 K \right)\int d\eta \cos\left(\eta\tilde z\right) \frac 1 {\eta} \sinh^2 r_\eta
\end{equation}
which contains the Fourier transform of the function $d^\prime(\eta)=\frac 1 {\eta} \sinh^2 r_\eta$, which is a function peaked around $\eta=1$ with height $\sinh^2\left(\frac{\pi n \delta}{4}\right)$ and width $\delta/2$. Fourier transforming it will give an oscillating part in the correlation function. \\

Directly calculating $\abs{ A_q}$ in the same way as for bosons does not give any peaks. In the next section we discuss why and how to resolve this issue. 
Using parametric driving and analyzing the momentum distribution of cold fermionic gases to study spin-charge separation has been proposed in \cite{kagan:023625, graf-epl-2009}. Here we propose to analyze an interference pattern. 

\section{Observing Spin Charge Separation for Fermions}
\label{SpinCharge}

In \fref{fig:2comp_bosons} we see two peaks in $\abs{A_q}^2$ for the two component bose system, corresponding to the sound velocities of two modes. 
In this section we show how one could observe spin charge separation for fermions in a similar way, however, for fermions there are some complications. 
The fast decay of correlation functions coming from the factor $(1/K+K)$ in the power law exponent poses a problem when calculating $\ave{\abs{A_q}^2}$, as this quantity will then be dominated by contributions from close to the origin $z=0$, preventing the Fourier transform from resolving the beats in the correlation function. 
We can solve this problem by using a simple trick to avoid the region around the origin (similarly as used in \cite{gritsev-2008-78}). We need to extract
\newcommand {\zbeg} {z^{\textrm{beg}}}
\newcommand {\zend} {z^{\textrm{end}}}
\begin{equation}
\int_{\zbeg}^{\zend} \cos(zq)\abs{\ave{\psidag(z)\hat \psi(0}}^2
\end{equation}
from the interference pattern, with $\zend>\zbeg>0$. 

\begin{figure*}[bt]
\begin{tabular}{l l}
a) & b)\\
\begin{minipage}{0.49\textwidth}
\includegraphics[width=\textwidth]{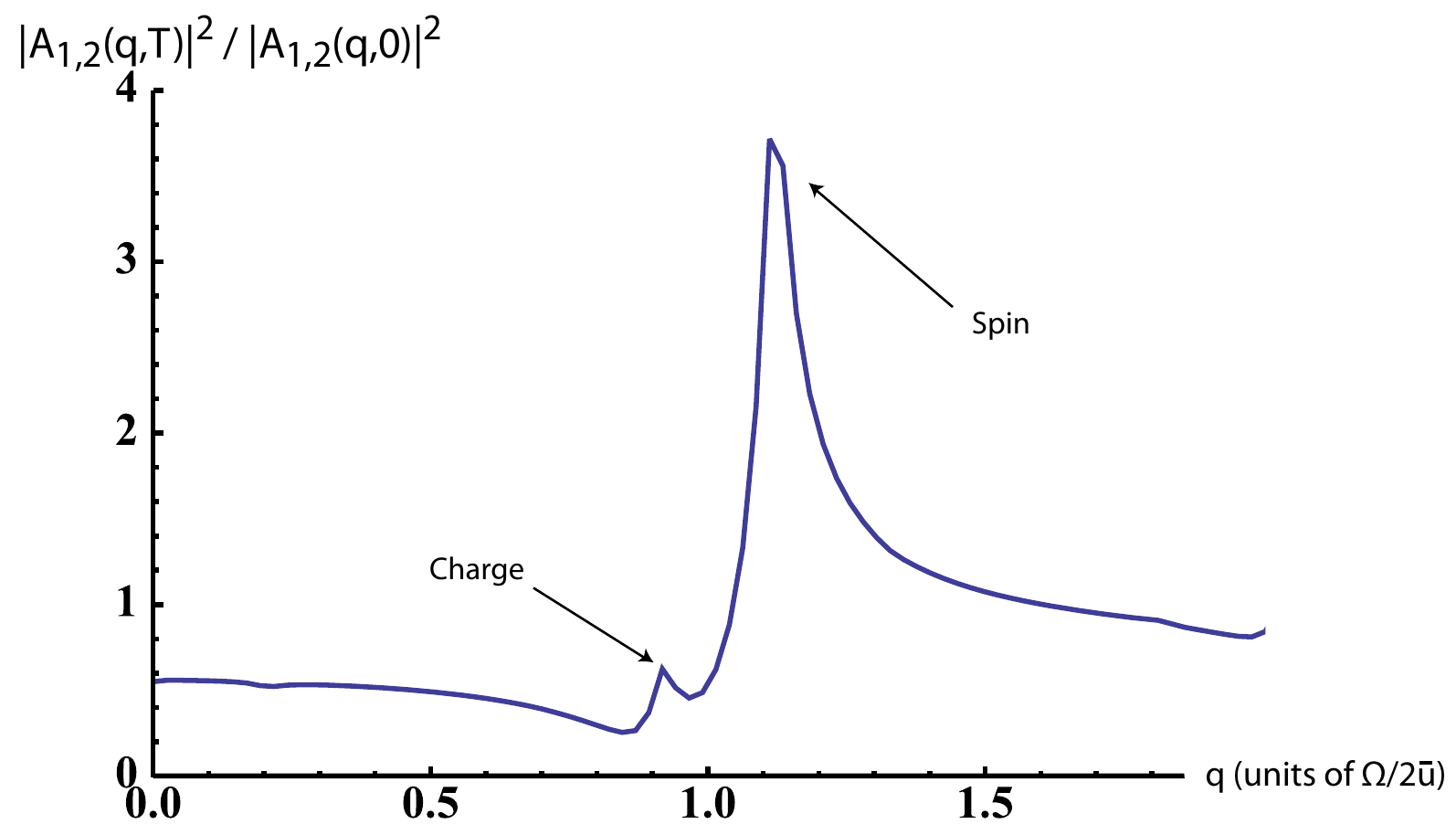}
\end{minipage}
&
\begin{minipage}{0.49\textwidth}
\includegraphics[width=.8\textwidth]{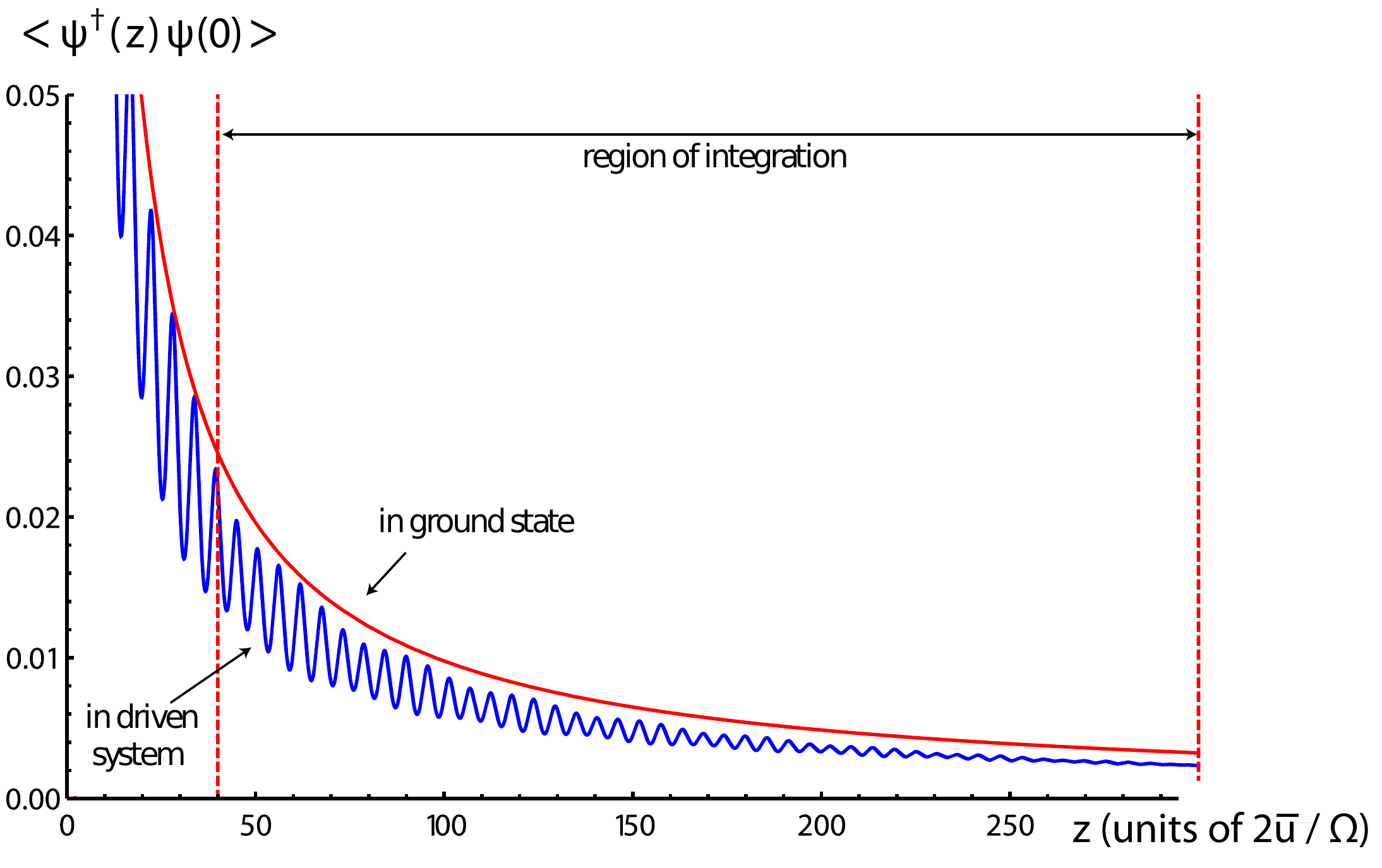}
\end{minipage}
\\
\end{tabular}
\caption{(Color online) 
Fermions with spin: (a) $\ave{\abs{A_{1,2}(q,T)}^2}$ shows a peak for the spin mode and one for the charge mode. To obtain these peaks, one has to integrate over different regions of the interference patter, see text. (b) Correlation function for driven two component system of fermions, and integration region. 
}
\label{fig:2comp_fermions}
\end{figure*}

This can be realized by looking at correlations of Fourier transformations of density of different regions in the interference pattern. For this purpose we define
\begin{equation}
\hat{\tilde\rho}_{1/2}(x,q)=\int_{\textrm{region 1/2}} e^{iqz} \hat \rho(x,z) dz
\end{equation} 
and choose the two regions so that they have no overlap. Then
\begin{equation}
\ave{\hat{\tilde\rho}_1(x, q)\hat{\tilde\rho}_2(0,-q)}= 2 \cos(xQ) \ave{A_\textrm{reg1}(q)A^\dag_\textrm{reg2} (q)}
\label{eq:Aq cut}
\end{equation}
where we defined $A_\textrm{reg i}(q)=\int_\textrm{region i} e^{iqz}\psidag_1(z)\hat\psi_2(z)dz$ and we have
\begin{equation}
\ave{A_\textrm{reg 1}(q)A^\dag_\textrm{reg 2}(q)}\approx L \int_{{\zbeg_2}-{\zend_1}}^{\zend_2-\zbeg_1}e^{izq}\abs{\ave{\psidag(z)\hat \psi(0}}^2.
\end{equation}
Note that because of the different regions the above quantity does not have to be real. We define
\begin{equation}
\abs{A_{1,2}(q)}^2=\abs{\ave{A_\textrm{reg 1}(q)A^\dag_\textrm{reg 2}(q)}}.
\end{equation}
This quanity as well as the correlation function is shown in \fref{fig:2comp_fermions}, where the red lines in the plot of the correlation function in mark $z_\textrm{begin}$ and $z_\textrm{end}$. $\abs{A_{1,2}(q)}^2$ shows two peaks corresponding to charge and spin velocities. The difference in size of the peaks is simply due to the different relative change in the sound velocities $u$, which are given by \eqnref{eqn:uFermi}, where the relative change in $g$ is of course equal for both modes. There are more excitations in the spin sector. 

\section{Conclusions}

We have studied interference of parametrically driven one-dimensional systems of  both bosonic and fermionic ultracold systems, and found that the interference pattern contains information about the excitations driven into the systems. 
For two component systems two kinds of excitations are driven on resonance: those corresponding to charge velocity, and those corresponding to spin velocity. This can be extracted from the interference pattern. 
In the case of fermions one needs to use different regions of the interference pattern to resolve the effect. 
We argued that such experiments could be used to probe properties of one-dimensional systems, as measuring the sound velocity, and spin charge separation. \\
\emph{Note added: } During the preparation of this manuscript the author became aware of Refs. \cite{kagan:023625, graf-epl-2009}, who studied similar effects for fermionic systems and obtained similar results. 

\begin{acknowledgments}
The author thanks E. Demler for heading her to this problem. I would like to thank S. Sachdev, R. Glauber, J. Hoffman, D. Pekker, and V. Gritsev for valuable discussions and feedback on this work. 
S.P. acknowledges support from the Studien\-stiftung des Deutschen Volkes.
\end{acknowledgments}

\begin{appendix}
\section{Derivation of equation 1}
\label{appendix1}
When we write down the creation operator $\hat \psi^\dag_{\rm cloud}(x_1,z_1)$ for a particle after expansion at position $(x_1, z_1)$, this particle can have come from either quasicondensate, and there is a relative phase between those paths 
\begin{equation}
\hat \psi^\dag_{\rm cloud}(x,z)\propto \psidag_1(z)e^{ix Q/2} + \psidag_2(z)e^{-ix Q/2} 
\end{equation}
where $\psidag_j(z)$, $j=1,2$, is the creation operator for a particle in quasicondensate $j$ before expansion, and we factored out a common phase. 
The relative phase comes from the expansion the particles initially confined in a harmonic potential (particles only occupy the ground state of the transversal confinement). Consider first a particle in a harmonic oscillator at position $x \pm d/2$, whose potential is suddenly removed. The wave function $f_\pm(x,t)$ expands and picks up a phase factor
\begin{equation*}
f_{\pm}(x,t)=
\frac{1}{\pi^{3/4} \sqrt{R_t}}
\exp \left[-
\frac{\left(x\pm\frac d 2\right)^2 \left(1-i\frac t {m R_0^2}\right)}
{2 R_t^2}
\right]
\end{equation*}
where $R_0=\sqrt{\frac \hbar {m\omega_\perp}}$ is the initial radius of the Gaussian wave function, $m$ is the particle mass, and $\omega_\perp$ is the harmonic oscillator potential, here the transverse confinement frequency, and 
\begin{equation}
R^2_t=R^2_0+\left(\frac{\hbar t}{m R_0}\right)^2.
\end{equation}
For large expansion times, so that $R_t\gg R_0$, $R_t\rightarrow \frac{\hbar t}{m R_0}$, and
\begin{equation}
f_{\pm}(x,t)\approx
\frac{1}{\pi^{3/4} \sqrt{R_t}}
\exp \left[i \frac {m} {2\hbar t }\left(x \pm \frac d 2\right)^2\right].
\end{equation}
As the two systems are initially independent, $\ave{\psidag_1(z)\psi_2(z)}=0$, and thus the expectation value
\begin{equation}
\ave{\psidag_{\rm cloud}(x,z)\psi_{\rm cloud}(x,z)}=\rm{const}.
\end{equation}
However, each single shot will have interference fringes due to interference of the different paths, which one can see in the expectation value of the density-density correlation function 
$\ave{\hat \rho(x_1,z_1)\hat\rho(x_2, z_2)}$
of the cloud. The operator for finding one particle at $r_1= (x_1,z_1)$, and the other one in $r_2=(x_2, z_2)$, is proportional to
\begin{eqnarray}
\hat\psi_{\rm cloud}(r_1)\hat\psi_{\rm cloud}(r_2)&\propto&\hat\psi_1(z_1)\hat\psi_1(z_2) e^{-i(x_1+x_2)Q/2}\nonumber\\
&&+\hat\psi_2(z_1)\hat\psi_2(z_2)e^{i(x_1+x_2)Q/2}\nonumber\\
&&+\psi_1(z_1)\hat\psi_2(z_2)e^{-i\Delta x Q/2}\nonumber\\
&&+\psi_2(z_1)\hat\psi_1(z_2)e^{i\Delta x Q/2}\nonumber\\
\end{eqnarray}
here $\Delta x = x_1-x_2$. The first two terms describe the case when the two particles come both from the same quasi condensate. The latter two terms describe the case when the two particles come each from a different quasi condensates: those are the terms giving an oscillating term in \eqnref{eq1} -- and the interference pattern. 
When we multiply the above with its complex conjugate, and take the expectation value, the following two terms give non-constant contributions: 
\begin{eqnarray}
&&\ave{\psidag_2(z_2)\psidag_1(z_1)\psi_2(z_1)\psi_1(z_2)} e^{i \Delta x Q}+\hc\nonumber\\
&=&\pm\ave{\psidag_2(z_2)\hat \psi_2(z_1)}\ave{\psidag_1(z_1)\hat \psi_1(z_2)}e^{i \Delta x Q}+\hc\nonumber\\
&=&\pm2\abs{\ave{\psidag(z_2)\hat \psi(z_1)}}^2\cos(\Delta x Q)
\end{eqnarray}
the upper (lower) sign is for bosons (fermions). In the first step we assumed that systems $1$ and $2$ are independent, thus the expectation value factorizes, in the second step we assumed that they are equal, i.e. in the same quantum state, and used $\psi$ to describe that state. 
\end{appendix}

\begingroup 
\interlinepenalty=10000 


\endgroup

\end{document}